# A study of centaur (54598) Bienor from multiple stellar occultations and rotational light curves

J. L. Rizos[1], E. Fernández-Valenzuela[2, 1], J. L. Ortiz[1], F. L. Rommel[3, 2, 4], B. Sicardy[5], N. Morales[1], P. Santos-Sanz[1], R. Leiva[1], M. Vara-Lubiano[1], R. Morales[1], M. Kretlow[1], A. Alvarez-Candal[1, 6], B. J. Holler[7], R. Duffard[1], J. M. Gómez-Limón[1], J. Desmars[8, 9], D. Souami[5, 10], M. Assafin[4, 11], G. Benedetti-Rossi[4, 5, 12], F. Braga-Ribas[3, 4, 13], J. I. B. Camargo[4, 13], F. Colas[12], J. Lecacheux[5], A. R. Gomes-Júnior[4, 12], R. Vieira-Martins[4, 5, 13], C. L. Pereira[4, 13], B. Morgado[4, 5, 13], Y. Kilic[14], S. Redfield[15], C. Soloff[15], K. McGregor[15], K. Green[16, 17], T. Midavaine[18], O. Schreurs[19, 20], M. Lecossois[20], R. Boninsegna[21], M. Ida[22], P. Le Cam[23], K. Isobe[24], Hayato Watanabe[25], S. Yuasa[25], Hikaru Watanabe[25], and S. Kidd[26]

*(Affiliations can be found after the references)*

Received May 22, 2024

## ABSTRACT

*Context.* Centaurs, distinguished by their volatile-rich compositions, play a pivotal role in understanding the formation and evolution of the early solar system, as they represent remnants of the primordial material that populated the outer regions. Stellar occultations offer a means to investigate their physical properties, including shape, rotational state, or the potential presence of satellites or rings.
*Aims.* This work aims to conduct a detailed study of the centaur (54598) Bienor through stellar occultations and rotational light curves from photometric data collected during recent years.
*Methods.* We successfully predicted three stellar occultations by Bienor, which were observed from Japan, Eastern Europe, and the USA. In addition, we organized observational campaigns from Spain to obtain rotational light curves. At the same time, we develop software to generate synthetic light curves from three-dimensional shape models, enabling us to validate the outcomes through computer simulations.
*Results.* We resolve Bienor's projected ellipse for December 26, 2022, determine a prograde sense of rotation, and confirm an asymmetric rotational light curve. We also retrieve the axes of its triaxial ellipsoid shape as a = (127 ± 5) km, b = (55 ± 4) km, and c = (45 ± 4) km. Moreover, we refine the rotation period to 9.1736 ± 0.0002 hours and determine a geometric albedo of (6.5 ± 0.5) %, higher than previously determined by other methods. Finally, by comparing our findings with previous results and simulated rotational light curves, we analyze whether an irregular or contact-binary shape, the presence of an additional element such as a satellite, or significant albedo variations on Bienor's surface, may be present.

## 1. Introduction

Among the small bodies of the solar system, there exists the so-called centaurs, some of them characterized by their combination of cometary and asteroidal features. It is theorized that centaurs are originally trans-Neptunian objects expelled due to gravitational scattering by Neptune (Holman & Wisdom 1993; Duncan et al. 1995; Duncan & Levison 1997). These objects, transient in nature, have an average lifespan of only a few million years (Horner et al. 2004), primarily because of their gravitational interactions with the giant planets that lead to their eventual expulsion from the region. Therefore, centaurs present a unique opportunity to study trans-Neptunian objects, providing a better characterization of their physical properties.

The first centaur discovered was Chiron in 1977, although its cometary properties were not identified until a decade later —to date, comet activity has been confirmed in ~13 % of cases (Bauer et al. 2008). Since the discovery of a second object in 1992, (5145) Pholus, hundreds[1] of them have been discovered, although the number depends on the criteria used for their definition. There is not a universally accepted definition of centaurs. Several criteria can be found in the literature, such as having a semi-major axis between 5 and 30 au or those presenting a Tisserand parameter larger than 3 and a semi-major axis larger than the semi-major axis of Jupiter. Some estimations suggest that there are > 40,000 centaurs larger than 1 km in diameter, although other authors have estimated > 100 million (Di Sisto & Brunini 2007).

In a study by Lacerda et al. (2014), more than 100 TNOs and centaurs were examined for their color-albedo distribution. The study identified two clusters: one characterized by dark-neutral objects with a geometric albedo ($p_V$) of approximately ~5%, and another composed of bright red objects with geometric albedos exceeding > 6%. This division based on color-albedo has been particularly pronounced among centaurs (Bauer et al. 2013; Duffard et al. 2014b; Tegler et al. 2016) with median albedos of approximately ~5% for the dark-neutral group and ~8.4% for the bright red group (Müller et al. 2020).

Due to the large geocentric distances of these small bodies and the consequent challenges in their study, we have only a limited understanding of the physical properties of these objects thus far. To date, through the technique of stellar occultations, we have gathered data on the shape and spatial orientation of the centaur Chariklo (Leiva et al. 2017; Morgado et al. 2021). Moreover, a three-dimensional morphology has been postulated for the centaur Chiron under the assumption of hydrostatic equilib-rium (Braga-Ribas et al. 2023). There are also constraints regarding the shapes of centaurs 2002GZ32 (Santos-Sanz et al. 2021; Strauss et al. 2021) and Echeclus (Rousselot et al. 2021; Pereira et al. 2024). In addition to their morphology, stellar occultations have revealed that at least two centaurs, Chariklo and Chiron, possess rings (Braga-Ribas et al. 2014; Ortiz et al. 2015, 2023).

(54598) Bienor ($q$ = 13.19 au, $Q$ = 19.90 au, $i$ = 20.74°) is an inactive centaur discovered by the Deep Ecliptic Survey in 2000 (Elliot et al. 2005), with an orbital period of 67.34 years and a rotation period of ~9.14 hours (Ortiz et al. 2002). Near-infrared spectral measurements indicated the presence of water ice on its surface (Dotto et al. 2003; Guilbert et al. 2009). A geometric albedo of (5.0 ± 1.9) % (Bauer et al. 2013), together with the fact that its B-V color is 1.12 ± 0.03 (Tegler et al. 2008), places Bienor in the dark-neutral group. Extreme light curve amplitudes (DeMeo et al. 2009) suggest Bienor as a highly elongated body. According to Fernández-Valenzuela et al. (2017), the orientation of its rotational pole is $\beta_p$ = 50° ± 3°, $\lambda_p$ = 35° ± 8° (prograde rotation) or $\beta_p$ = −50° ± 3°, $\lambda_p$ = 215° ± 8° (retrograde rotation). Assuming Bienor is a triaxial ellipsoid under hydrostatic equilibrium, they constrained the $b/a$ axial ratio to 0.45 ± 0.05 and a density of 594 $kg \cdot m^{-3}$, although other solutions are also feasible.

Some controversy surrounds this object due to the discrepancy in size and albedo obtained through different techniques (Lellouch et al. 2017). Through thermal measurements, Bauer et al. (2013), using data from the Wide-field Infrared Survey Explorer (WISE), obtained (187.5 ± 15.5) km of diameter and a geometric albedo of (5.0 ± 1.9) %. Duffard et al. (2014a) retrieved $198^{+6}_{-7}$ km of diameter and a geometric albedo of $4.3^{+1.6}_{-1.2}$ using radiometric measurements from Herschel Space Observatory and Spitzer Space Telescope, and Lellouch et al. (2017) obtained from Spitzer, Herschel and ALMA observations, a diameter of 179–184 ± 6 km and a geometric albedo of 5.0–5.3 $^{+1.8}_{-1.6}$. However, Fernández-Valenzuela et al. (2023) estimated an area-equivalent diameter of 150 ± 20 km using the stellar occultation technique. This is smaller than the thermal diameters and suggestive of a possible satellite or binarity. Moreover, the authors reported a strong irregularity in one of the minima of the rotational light curve regardless of the aspect angle, but no features compatible with rings or satellites within their data. Nevertheless, they do not discard a satellite or a ring system similar to that of the centaurs Chariklo and Chiron, due to the observations' low signal-to-noise ratio (S/N).

These discrepancies and irregularities have sparked a question about the possible presence of a ring, satellite, or the possibility that it is a binary system. The objective of this study is to delve further into this issue. For that, as part of our global collaboration to detect stellar occultations by outer Solar System bodies, we successfully predicted and observed three positive occultations that took place on February 6 and December 26, 2022, and on February 14, 2023. Alongside this data, we also conducted several observational campaigns to obtain two new rotational light curves that complement the occultation data. In this work, we present our observations and their corresponding technical details in Section 2, followed by the data analysis outcomes in Section 3. Section 4 introduces our software for simulating synthetic rotational light curves, and then it is applied to contextualize the preceding results. Finally, in Section 5 we discuss our findings and check various hypotheses that would explain the discrepancies with the results obtained by other authors through different methods, with Section 6 summarizing the conclusions drawn.

## 2. Observations

### 2.1. Stellar occultations

Primary predictions of stellar occultations by Bienor were performed using the Numerical Integration of the Motion of an Asteroid (NIMA) solution (Desmars et al. 2015). Subsequently, after a series of observational campaigns carried out during the last years (see Fernández-Valenzuela et al. 2023), Bienor's orbit was updated and refined. This has allowed the prediction and observation of three positive occultations in 2022 and 2023 described in the following sections.

#### 2.1.1. Occultation A - February 6, 2022

On February 6, 2022, an occultation by Bienor of the star Gaia DR3 202357903847210880 (see Table 1) was predicted to happen at 08:52:54 UT (referred hereafter as "Occ. A"), with a maximum duration of 14.7 seconds in the centrality.

The event was observed from two different locations, resulting in positive detections from Middletown and Westport (Connecticut, USA), as shown in Fig. 1a. Table B.1 in Appendix A expands on the information regarding location, instrumentation, and observers involved. Time-series observations were obtained and time-synchronized using Network Time Protocol (NTP) servers or Global Positioning Systems (GPS) devices. Image acquisition started at least ~5 min before and ended ~5 min after the predicted time. No filters were used to maximize the S/N of the occulted star.

The diameter of the occulted star was estimated using the following equation (van Belle 1999):

$$\theta = \frac{10^{a+b(V-K)}}{10^{V/5}}. \tag{1}$$

with $(a, b) = (0.5, 0.264)$ for main-sequence stars, or $(0.669, 0.223)$ for giant and supergiant stars, and where $V$ and $K$ denote the apparent magnitudes of the star in the V-band and K-band, respectively. According to Zacharias et al. (2004), $V$ = 12.93 mag and $K$ = 10.36 mag. Consequently, because the occulted star is a giant (see Fig. A.1 in Appendix), the angular diameter is computed as 0.0453 milliarcseconds (mas), corresponding to 0.44 kilometers at a Bienor's geocentric distance ($\Delta$) of 13.39 au. Because $\lambda$ is ~600 nm for our CCD/CMOS observations, from Eq. 2, the Fresnel scale of this object is calculated as 0.78 kilometers.

$$F = \sqrt{\frac{\lambda \Delta}{2}} \tag{2}$$

The minimum cycle time (exposure plus dead time) here comes from the observation made in Westport, which is recorded as 0.25 seconds. With Bienor's velocity at the time of occultation measured at 12.81 km/h, this cycle time corresponds to a distance of 3.20 km. Given that this distance is an order of magnitude greater than that derived from the star's angular size and Fresnel diffraction effects, they have a small impact on the derivation of ingress and egress times.

#### 2.1.2. Occultation B - December 26, 2022

On December 26, 2022, an occultation by Bienor of the star Gaia DR3 961231167113907328 (Table 1) with a maximum duration of 8.6 seconds was predicted to happen at 20:13:44 UT (referred hereafter as "Occ. B"). The star was observed from seven locations, all reporting positive detections (Fig. 1 b and c). The telescopes were in Eastern Europe and Japan (see Appendix A Table B.1 for more information). Time-series observations were obtained and synchronized using GPS devices. Image acquisition started at least ~5 min before and ended ~5 min after the predicted time. No filters were used to maximize the S/N of the star.

Table 1: Occulted stars by Bienor presented in this work. Stellar right ascension and declination represent the astrometric coordinates with proper motion based on the ICRF.

| Occultation | Source | Right Ascension (ICRF) | Declination (ICRF) | G, V, K Magnitudes |
|---|---|---|---|---|
| A | 202357903847210880 (GaiaDR3) | 05 12 37.5915 | +43 49 59.539 | 12.68, 12.93, 10.36 |
| B | 961233167113907328 (GaiaDR3) | 06 06 16.6854 | +43 59 24.029 | 14.43, 13.98, 12.87 |
| C | 193308850132660224 (GaiaDR3) | 05 54 29.1042 | +43 13 29.376 | 16.86, 16.45, 15.26 |

A diameter of 0.0132 mas of the occulted star was estimated using Eq. 1 (giant star, see Fig. A.1 in Appendix), with $V = 13.98$ mag and $K = 12.87$ mag from Zacharias et al. (2004). It corresponds to 0.12 kilometers at Bienor's geocentric distance, $\Delta$, of 12.81 au. The Fresnel scale in this case is 0.76 km. Because the minimum cycle time for these observations is 0.16 seconds, while Bienor's velocity at the time of occultation was 21.76 m/h, it translates to a distance of 3.48 km. Thus, the main source of uncertainty comes from the ingress and egress times of the photometric data.

### 2.1.3. Occultation C - February 14, 2023

On February 14, 2023, an occultation by Bienor of the star Gaia DR3 193308850132660224 (Table 1) with a maximum duration of 14.6 seconds was predicted at 01:43:46 UT (referred to hereafter as "Occ. C"). The star was observed from two telescopes, resulting in both cases in positive detections. The telescopes were located in Belgium and England (see Fig. 1d and Table B.1). Time-series observations were obtained and synchronized using GPS devices. Image acquisition started at least ~5 min before and ended ~5 min later than the predicted time. No filters were used to maximize the S/N of the star.

A diameter of 0.0033 mas of the occulted star was estimated using Eq. 1 (main-sequence star, see Fig. A.1 in Appendix), with $V = 16.45$ mag and $K = 15.26$ mag from Zacharias et al. (2004). This corresponds to 0.03 kilometers at Bienor's geocentric distance, $\Delta$, of 13.16 au. The Fresnel scale in this case is 0.77 km. Because the minimum cycle time for these observations is 1.603 seconds, while Bienor's velocity at the time of occultation was 12.87 m/h, it translates to a distance of 20.6 km. Given that this distance is orders of magnitude greater than that derived from the star's angular size and Fresnel diffraction effects, they have a small impact on the derivation of ingress and egress times.

### 2.2. Rotational light curves

To determine the rotational phase of Bienor at the moment of the occultations, photometric data were collected in 2021 and 2023 using the 1.5 m telescope at the Sierra Nevada Observatory (OSN) and the 1.23 m telescope at the Calar Alto Observatory (CAHA). The OSN 1.5 m telescope has a 2k×2k CCD with a field of view (FoV) of $7.92' \times 7.92'$ and a pixel scale of $0.232''$/pixel. The CAHA 1.23 m telescope has a 4k×4k CCD with a FoV of $21.5' \times 21.5'$ and a pixel scale of $0.315''$/pixel. In both cases, binning 2×2 was used.

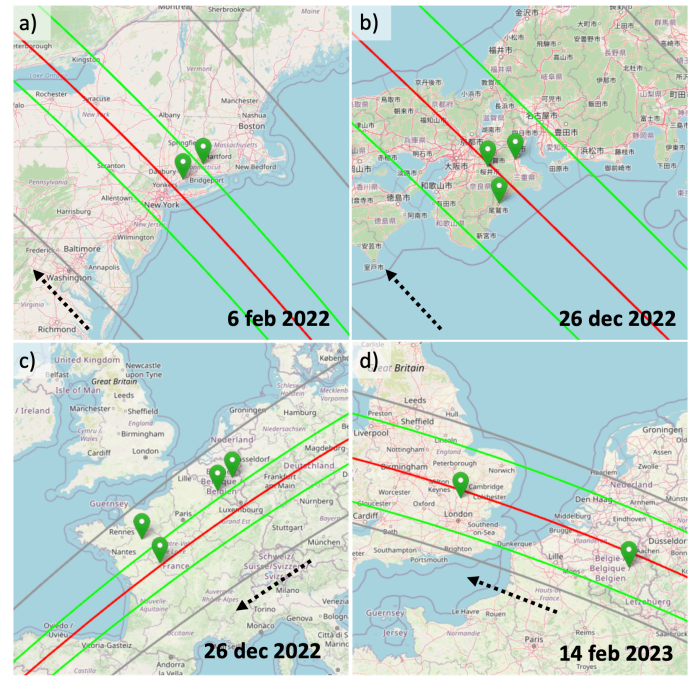

Fig. 1: Predicted occultation shadow path computed for February 6, 2022 (a), December 26, 2022 (b & c), and February 14, 2023 (d). Green lines depict the boundaries of the shadow paths, while the gray lines represent the uncertainty in the path resulting from orbital uncertainties. A red line denotes the center of the shadow path, and the pins indicate the positions of the stations involved in the campaigns. Green pins signify sites where a positive occultation was observed. The maps were generated using OpenStreetMap via Occultation Portal. The dotted arrow denotes the shadow's direction of motion.

Observations of Bienor were conducted at OSN on October 8, 9, 10, 16, 18, 19, and 20, 2021, followed by another set of observations on December 12 and 13, 2023. At CAHA, observations were carried out on October 11, 12, and 21, 2021, and November 14 and 15, 2023. An exposure time of 300 s and an R Johnson filter was used in all the observations made with these telescopes.

## 3. Data analysis and results

### 3.1. Stellar occultations

Data were compiled and managed through the Tubitak Occultation Portal website (Kilic et al. 2022). Firstly, images were bias and flat-field calibrated using standard procedures with the AstroImageJ software (AIJ) (Collins et al. 2017). For those cases in which the observations were recorded in video format, we used the Planetary Imaging PreProcessor (PIPP[2]) software to convert them to FITS before performing the photometry.

Next, we performed time-series multi-aperture differential photometry with AIJ, measuring the flux of the occulted star (blended with Bienor) and dividing by the comparison stars selected in the FoV. This allows the minimization of systematic photometric errors due to atmospheric variability. As the event only lasts several seconds, flux variations due to the rotational variability of the body do not affect the resulting light curve. We chose an aperture and sky-background inner and outer radius to minimize the noise of the data outside the main flux drop due to the occultation. We derived the error bars from Poisson noise calculations and the results were scaled to the standard deviation of the data (see for details Ortiz et al. 2020). Finally, ingress and egress timing and their uncertainties are computed using the Python Occultation Timing Extractor (PyOTE[3], v5.5.1) software package.

The light curves obtained are shown in Appendix A (Fig. B.1,B.2,B.3,B.4). Table B.2 shows the extracted ingress and egress times along with their respective uncertainties.

To project the occultation chords in the sky-plane, we need to translate the ingress and egress time values by using the astrometric right ascension and declination of Bienor with respect to the observing sites (ICRF frame, JPL#71 ephemerids). For that, we use the JPL Horizons online solar system data and ephemeris computation service[4]. We then subtract the value of these coordinates (extremities values of the chords) from the value of the occulted stars' coordinates, and convert this difference into kilometers in the sky-plane defined by Bienor at its geocentric distance. The result for each occultation is shown in Fig. 2a, 2b and 2c. All plots present the same vertical and horizontal scale (320 × 320 km) for better comparison.

Because the general equation of an ellipse (conic section) has five degrees of freedom –two for the position of the center $(x_0, y_0)$, two for the lengths of the semi-major and semi-minor axes $(u, v)$, and one for the rotation of the ellipse or position angle–, we need at least 5 points to find a unique solution; otherwise, the solution degenerates. Thus, we can only fit an ellipse to the extreme of the chords obtained on December 26, 2022. For that, we used the numerically stable and non-iterative least squares algorithm defined by Halíř & Flusser (1998). The resulting ellipse along with the chords is shown in Fig. 2b. To propagate ingress and egress time uncertainties into this fitted ellipse, we run a Monte-Carlo algorithm by generating 1000 random clone extreme points following a Gaussian distribution centered at the value of the chord endpoint and with a width equal to the error in each node. For each cloned solution, we fit an ellipse, so that the final parameters of the ellipse (Table 2) are determined by the average value obtained from the 1000 solutions, with the error represented by the standard deviation. The choice of 1000 clones is because we empirically confirm that at this value, the parameters have reached convergence, and further increasing the number of clones does not enhance the precision of the outcome.

The result is that Bienor is a highly elongated triaxial ellipsoid, which is compatible with the range of solutions given in Fernández-Valenzuela et al. (2017).

Finally, using the rotation pole solutions given by Fernández-Valenzuela et al. (2017) and Bienor ephemerides at each occultation time, we compute both the aspect angle and the position

Table 2: Parameters of the fitted ellipse for the chords obtained on December 26, 2022.

| | |
|---|---|
| Center coordinates $(x_0, y_0)$ (km) | (15.5 ± 2.7, 12.9 ± 3.8) |
| Semi-major axis, u (km) | 107.0 ± 3.7 |
| Semi-minor axis, v (km) | 51.9 ± 2.1 |
| Position angle (°) | 19.3 ± 3.4 |

Table 3: Bienor aspect angle ($\psi$) and position angle of the rotation pole ($\theta_p$), for prograde (1) and retrograde (2) solutions.

| | Occ. A | Occ. B | Occ. C |
|---|---|---|---|
| $\psi_1$ (°) | 133.63 | 127.14 | 127.98 |
| $\theta_{p1}$ (°) | 315.58 | 318.69 | 318.18 |
| $\psi_2$ (°) | 46.37 | 52.86 | 52.02 |
| $\theta_{p2}$ (°) | 135.58 | 138.59 | 138.18 |

angle of the rotation pole (prograde and retrograde). This result is shown in Table 3.

## 3.2. Rotational light curves

We corrected the images from bias and flat-field using the same tools and standard procedures described in Section 3.1. Subsequently, we performed time-series multi-aperture differential photometry by selecting the same reference stars set within each observation run.

To compute the rotational phase, $\phi$, we applied Eq. 3:

$$\phi = \frac{JD - JD_0}{P} \qquad (3)$$

where $JD$ is the Julian date of each image, $JD_0$ is an arbitrary initial date, and $P$ is the sidereal rotational period. All times in this equation are corrected for light travel time. Note that in this work the rotational phase is computed such that it corresponds to zero when the light curve reaches maximum brightness.

Fernández-Valenzuela et al. (2017) determined a synodic period for Bienor of 9.1713 ± 0.0002 h. However, this value does not fit our data. For example, using 2021 photometric data, we determine a rotational phase value of Bienor during the Occ. B of 0.42 ± 0.02, while using 2023 data, the phase value was 0.96 ± 0.02.

We first suspected that it could be due to the difference between synodic and sidereal periods. As discussed by Harris et al. (1984), the Phase Angle Bisector (PAB), i.e., the line that bisects the heliocentric and geocentric direction to the small body, serves as an approximation to estimate the difference between sidereal and synodic periods. It can be determined by applying Eq. 4:

$$|P_{syn} - P_{sid}| \sim \frac{\Delta p}{2\pi} \frac{P_{syn}^2}{\Delta t} \qquad (4)$$

where $\Delta p$ represents the longitude variation of the phase angle bisector and $\Delta t$ is the time span of observations to derive $P_{syn}$. This equation is invalid when the small body has a nearly polar aspect, but here it is not the case (see Table 3).

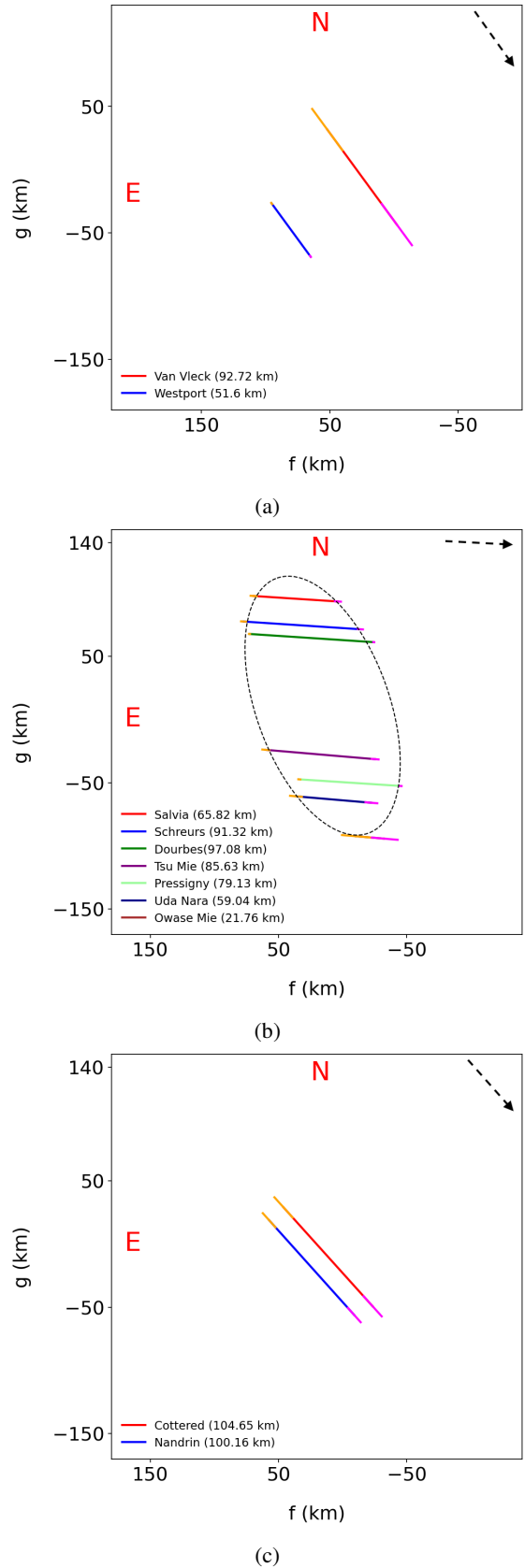

Fig. 2: Chords of the stellar occultations projected on the sky plane. The ingress uncertainties are shown in pink color and the egress uncertainties are in orange. The values in parenthesis are the lengths of the chords. (a) Occ. A on February 6, 2022. The minimum distance between the chords is 71.49 km. (b) Occ. B on December 26, 2022. The dashed line describes the ellipse that best fits the points. (c) Occ. C on February 14, 2023. The minimum distance between the chords is 15.35 km. The dashed arrow in the upper right corner indicates the direction of the shadow motion.

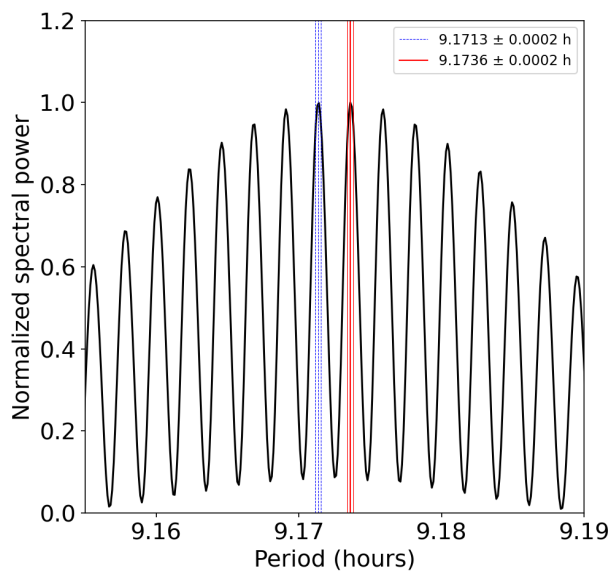

Fig. 3: Lomb periodogram's spectral power obtained from photometric data collected in 2021 and 2023. There are two possible solutions that are very close in spectral power. The solution that best fits with the 2021 and 2023 rotational light curves is the one marked in red, (9.1736 ± 0.0002) h.

We compute the PBA longitude variation during the period of time in which the observations to derive the synodic period were made, i.e., from 6/12/2013 to 8/8/2016. During this time range, $\Delta p$ was 18.93°, and according to Eq. 4, $|P_{syn} - P_{sid}| \sim$ 0.00018 h, the same order of magnitude than the error associated with the synodic rotational period, insufficient value to explain the discrepancy.

Then, we employed the Lomb algorithm (Lomb 1976), a variant of Fourier spectral analysis as coded by Press et al. (1992), to identify the rotation period with our new photometric data. The resulting periodogram is depicted in Fig. 3, wherein the best-fitting solution corresponds to a rotation period of (9.1736 ± 0.0002) h (red line). We see that another very close solution appears with virtually the same spectral power, which coincides with the solution given by Fernández-Valenzuela et al. (2017) (blue line). Thus explaining the discrepancy.

Finally, we folded the data so that the rotational light curves start at the absolute maximum, and fit the data points to a third-order Fourier function as follows:

$$m(\phi) = \sum_{n=0}^{i} a_n \cos(n\pi\phi) + a_n \sin(n\pi\phi), \qquad (5)$$

with $i = 3$.

The Fourier coefficients are presented in Table 4. Figures 4a and 4b display the rotational light curves for 2021 and 2023, respectively. Vertical lines within each plot denote the rotational phase for individual occultations. Dashed gray lines represent the uncertainty resulting from the propagation of a 0.0002-hour error over time. At the bottom of each plot, the residual difference between the modeled curve and observational data is depicted. To calculate the rotational phase of Occ. A, we utilized the 2021 rotational light curve, while for Occ. B and C, we relied on the 2023 light curve. This choice is based on the proximity of these datasets in time, resulting in lower error propagation. Specifically, Occ. A occurred 108 days after collecting the 2021 data. In contrast, Occ. B and C are separated by 323 and 275 days, respectively, from the date of the 2023 data collection. Table 5 shows the rotational phase values along with the propagated period error.

Table 4: Fourier coefficients after applying Eq. 5 to rotational light curves from 2021 and 2023.

| Fourier coefficients | 2021 RLC | 2023 RLC |
|---|---|---|
| $a_0$ | -0.010 | 0.008 |
| $a_1$ | -0.027 | -0.009 |
| $b_1$ | 0.083 | 0.067 |
| $a_2$ | -0.122 | -0.224 |
| $b_2$ | -0.037 | -0.034 |
| $a_3$ | -0.016 | -0.007 |
| $b_3$ | -0.004 | -0.004 |

Table 5: Rotational phase values of the occultations analyzed in this work. The error bars indicate the propagation of the period error.

| 6 Feb. 2022 (A) | 26 Dec. 2022 (B) | 14 Feb. 2024 (C) |
|---|---|---|
| 0.888 ± 0.006 | 0.15 ± 0.02 | 0.93 ± 0.02 |

As noted by Fernández-Valenzuela et al. (2023), the asymmetry in the minima of the rotational light curves is still present, given that in both curves the absolute minimum presents a ~45% deeper drop than the other one. Moreover, the light-curve amplitude ($\Delta m$) increased from 0.23 ± 0.01 in 2021 to 0.30 ± 0.01 in 2023.

### 3.3. Geometric albedo

Using the projected shape of Bienor during Occ. B, we retrieve the area-equivalent diameter ($D$), which is (149 ± 4) km. On that date, the value of Bienor's average visual magnitude ($H_V$) was 7.60 ± 0.04 (Fernández-Valenzuela et al. 2017). From the rotational light curve (Fig. 4b), at the time of the occultation, Bienor's magnitude was 0.12 units higher than the average, corresponding to an instantaneous $H_V$ of 7.72 ± 0.04. Using these values in Eq. 6 (Russell 1916):

$$\sqrt{p_V} = \frac{C}{D} 10^{-H_V/5} \qquad (6)$$

where $C = 1330 \lesssim 18$ km is a constant (Masiero et al. 2021), we determine $p_V = 6.5 \pm 0.5\%$, larger than the 5.0 ± 1.9%, $4.3^{+1.6}_{-1.2}$ % or $5.0–5.3^{+1.8}_{-1.6}$ % of previous thermal estimates from Bauer et al. (2013); Duffard et al. (2014a); Lellouch et al. (2017), respectively.

### 3.4. 3D shape and sense of rotation

From Magnusson (1986), the mathematical relationship between the major and minor axes of the projected ellipse ($u, v$), aspect angle ($\psi$), rotational phase ($\phi$), and the absolute value of the axes of the ellipsoid $a$, $b$ and $c$, is given by:

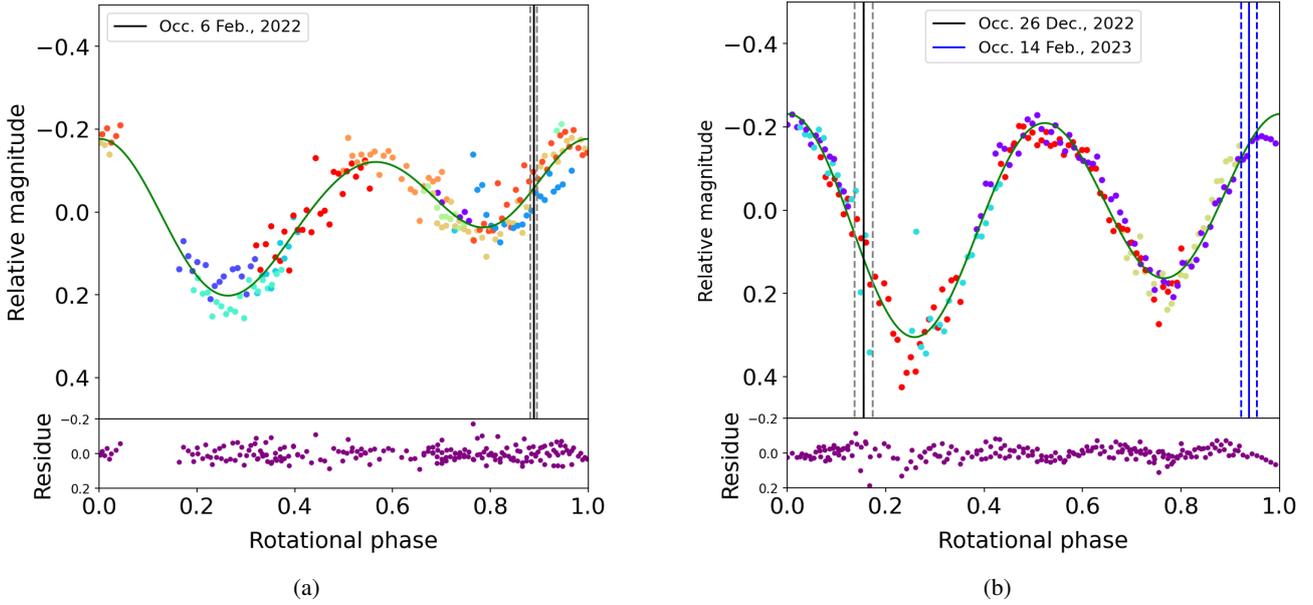

(a)

(b)

Fig. 4: (a) Rotational light curve using 2021 observational data. The vertical line indicates the rotational phase during Occ. A on 6 Feb. 2022. (b) Rotational light curve using 2023 observational data. The vertical lines indicate the rotational phase during Occ. B and C on 22 Dec. 2022 and 14 Feb. 2023, respectively. In both panels: Dashed lines show the uncertainty after studying how the 0.0002 h error propagates with time. A unique color represents each observation day. At the bottom, it is shown the difference (residual) between the modeled curve and the observational data.

$$u = \left( \frac{2A}{-B - \sqrt{B^2 - 4A}} \right)^{1/2} \qquad v = \left( \frac{2A}{-B + \sqrt{B^2 - 4A}} \right)^{1/2} \qquad (7)$$

where:

$$A = b^2 c^2 \sin^2 \psi \sin^2 \phi + a^2 c^2 \sin^2 \psi \cos^2 \phi + a^2 b^2 \cos^2 \psi$$

$$-B = a^2 (\cos^2 \psi \sin^2 \phi + \cos^2 \phi) + b^2 (\cos^2 \psi \cos^2 \phi + \sin^2 \phi) + c^2 \sin^2 \psi$$

From $u$, $v$, $\psi$, and $\phi$ derived from Occ. B, we conducted a $\chi^2$ minimization based on a grid search for the absolute axes $a$, $b$, and $c$ through the Eq. 7. We constrain the search so $\Delta m$ is within the light curve range determined in 2021 and 2023, that is, 0.23 and 0.30 mag, and then we explored all feasible solutions of $u$ and $v$ ensuring that $\chi^2$, which consider both the values and errors of $u$ and $v$, as well as the elongation ($b/a$) and flattening ($c/b$) metrics as described by Fernández-Valenzuela et al. (2017), is minimized. We find that the solution that best fits all the parameters is given by $a$=(127 ± 5) km, $b$=(55 ± 4) km, and $c$=(45 ± 4) km.

Assuming this solution, we calculate the projected area of Bienor during the date of the observations that led Lellouch et al. (2017) to determine an effective area of 179–184 ± 6 km. These observations were conducted using both Herschel and ALMA in 2011 and 2016. The aspect angles of Bienor on these dates were 140.42° and 147.05°, respectively, being 143.73° the intermediate aspect angle. Then, we determine the average value of the projected ellipse axes for a complete solution, resulting in $u$ = (116.73 ± 4.41) km and $v$ = (53.18 ± 3.59) km. This gives us a Bienor area-equivalent diameter of (158 ± 16) km, which is lower than the aforementioned estimation.

Finally, the position angle of the minor axis of the projected ellipse relative to the rotation pole ($\gamma$), can be computed using the following equation (Magnusson 1986):

$$\gamma = \frac{1}{2} \tan^{-1} \left( \frac{2 \cos \psi \, \cos \phi \, \sin \phi \left( \frac{b^2}{a^2} - 1 \right)}{\cos^2 \psi \, \sin^2 \phi - \cos^2 \phi + \frac{b^2}{a^2} \left( \cos^2 \psi \, \cos^2 \phi - \sin^2 \phi \right) + \frac{c^2}{a^2} \sin^2 \psi} \right).$$
(8)

And from $\gamma$, we determine the position angle of the minor axis of the projected ellipse with respect to the equatorial north, $\theta_v$, which is given by:

$$\theta_v = \theta_p + \gamma.$$ (9)

We obtain $\theta_v = 280.5°$ and $\theta_v = 176.68°$ for prograde and retrograde solutions, respectively. In Fig. 5 we plot both solutions (grey arrows) and compare them to the minor axis of the fitted ellipse (denoted by a dashed line). As can be seen, only the prograde rotation is compatible with the Occ. B fitted ellipse. Therefore, we determine that the sense of rotation of Bienor is prograde, excluding the retrograde solution, this being the first time that the rotational sense of a centaur has been determined.

## 4. Synthetic rotational light curves

The application of light curve inversion for determining the shape of small bodies has primarily been confined to main belt asteroids (Ostro et al. 1988; Durech et al. 2010). However, this method faces limitations when applied to centaurs or trans-Neptunian objects, owing to the need for observing these bodies from multiple angles and maintaining consistent sampling. These requirements pose significant challenges due to the considerable distance, small size, and slow movement across the sky of such objects, rendering the approach unfeasible (e.g. Showalter et al. 2021).

In this study, we employ an alternative technique known as forward modeling. Here, we construct a shape model based on

(a)

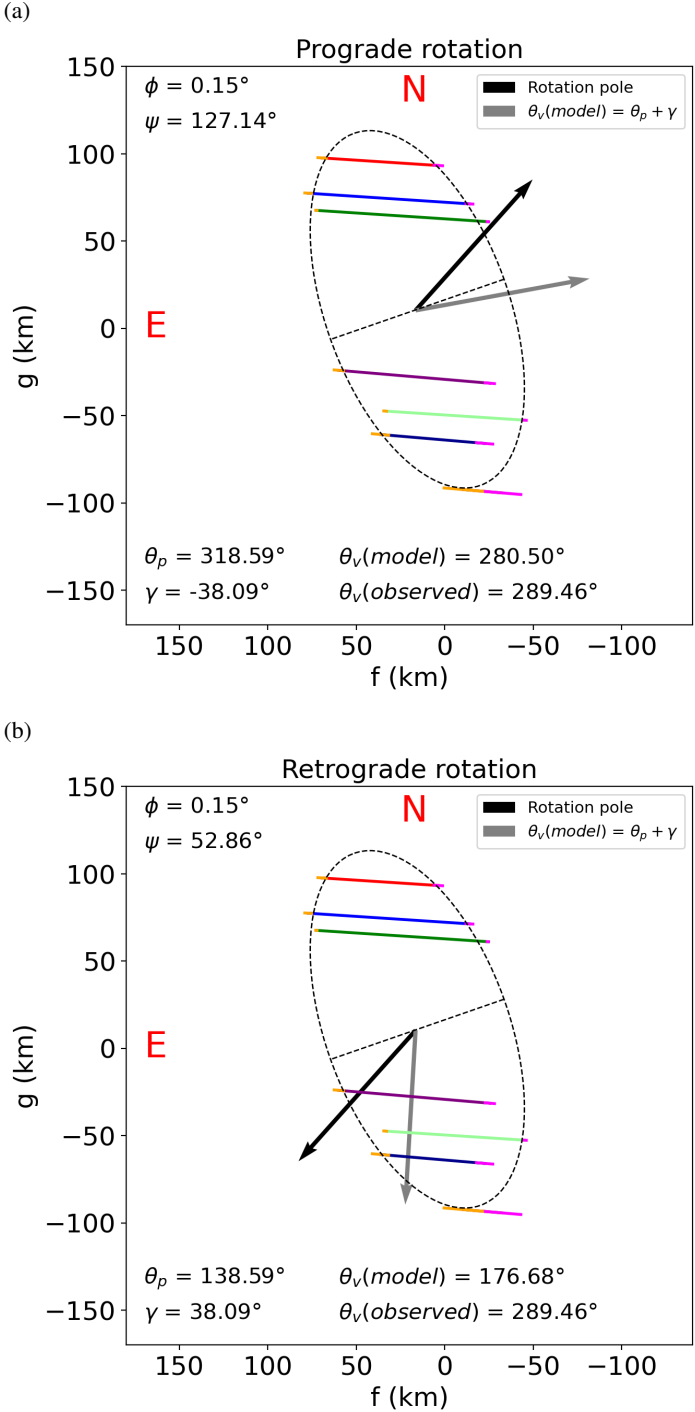

Fig. 5: a) Chords of Occ. B, the fitted ellipse, and the minor axis of the ellipse determined from Eq. 8 and 9, assuming a prograde rotation. b) Same for a retrograde rotation. Only the prograde solution is compatible with the fitted ellipse.

the known axis ratios of Bienor and the shape inferred from the previous occultations outlined in this study. Then, we simulate its rotation as it would be observed from Earth for each occultation date. While we acknowledge that a light curve does not yield a unique interpretation, the available dataset is sufficient to address the issue from this perspective. The main goal of this software is not about obtaining numerical results, but rather combining all the results obtained through different approaches to check whether they are compatible or not.

Initially, we constructed a triaxial ellipsoid shape model with semi-axis $a > b > c$ where $a = 1$, and $b$ and $c$ are given from the ratios provided by Fernández-Valenzuela et al. (2017), i.e., $b/a = 0.45$ and $c/b = 0.79$. This is done through Blender[5] software, which offers the capability to execute Python scripts. The resulting ellipsoidal shape model comprises a triangular mesh consisting of 920 facets, which provides adequate spatial resolution without excessively extending the calculation time.

Next, through JPL Horizons, we obtain the ecliptic Cartesian coordinates of Bienor, the Earth, and the Sun for each respective date. We wrote a Python-based software using the Poly3DCollection package of Matplotlib (Hunter 2007) to orient the shape model such that its rotation pole aligns with the specified values of $\beta_p = 50°$ and $\lambda_p = 35°$ (Fernández-Valenzuela et al. 2017). This allows us to generate an image of the body as seen from Earth and illuminated by the Sun according to the relative positional values. An illustrative example is shown in Fig. 6, where Bienor is depicted alongside its rotation pole denoted by the red arrow, the Earth and its rotation pole are in blue, and the Sun is represented in orange. The distance between the Earth and the Sun is not to scale for better visualization. The XYZ frame denotes the ecliptic Cartesian coordinates centered on the shape model center. The color code in the shape model corresponds to the solar incidence values, indicating the angle between the normal of each facet and a vector directed towards the Sun.

From the positions of the three bodies, we generated photometric backplanes to calculate the incidence, emission, and phase values for each facet. Then, we determine the radiance, that represents the radiant flux reflected per unit solid angle and unit projected area. This calculation assumes a Lambertian surface, meaning that the radiance remains constant regardless of the angle from which it is viewed, resulting in isotropic radiation. Finally, the software generates 2-D images of the ellipsoid in the sky-plane, with the north pointing towards the top of the FoV and the east towards the left, as it would be seen from Earth. These images allow visualizing the evolution of the small body at different temporal scales.

Therefore, we generate synthetic rotational light curves at specific epochs for comparison with the observations of Bienor made over the last 22 years.

### 4.1. Comparison between the observed and the synthetic rotational light curves

Thanks to the extensive research conducted by our group over the past two decades (Ortiz et al. 2002; Fernández-Valenzuela et al. 2017, 2023), a total of seven rotational light curves spanning from 2001 to 2023 have been compiled. To assess again the accuracy of the pole solution proposed by Fernández-Valenzuela et al. (2017) and to compare it with our synthetic light curve generation tool, we computed, for each date, the rotational light curve that our ellipsoid shape model would produce as observed from Earth with a prograde rotation period of 9.1736 h. The results are shown in Fig. D.2 in Appendix C. Our simulations align with the observed amplitude variations over this 22-year period, confirming the compatibility of the pole solution with the observations. However, it's worth noting that the observed light curves exhibit significant asymmetry, which becomes more pronounced as the amplitude increases. These asymmetries cannot be replicated by a symmetrical triaxial shape model.

---

[5] https://www.blender.org

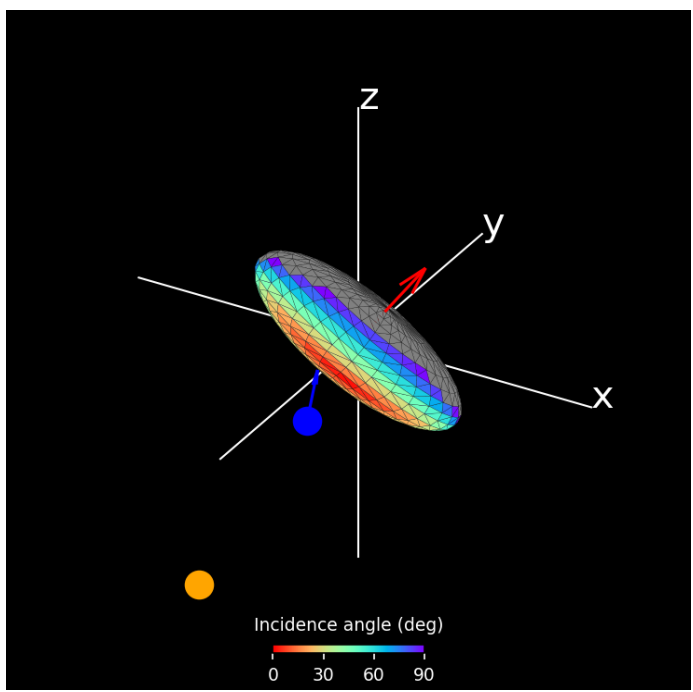

Fig. 6: Configuration of the Sun, the Earth, and Bienor on December 26, 2022, at the moment of Occ. B as generated by our software. Bienor is positioned at the center of the reference system in ecliptic coordinates, with its rotation pole indicated by the red arrow. The Earth is represented by the blue sphere, with its rotation pole marked by the blue arrow. The Sun is depicted as a yellow sphere. Geocentric and heliocentric distances are not to scale for better visualization. In the modeled Bienor, the color code corresponds to the values of solar incidence, representing the angle between the normal of each facet and a vector directed towards the center of the Sun. Gray facets are regions not illuminated by the Sun.

### 4.2. Overlaying chords on simulated projected ellipses

After validating the software and the rotation pole solution, we proceeded to conduct a detailed examination of the occultation on December 26, 2022, labeled as Occ. B.

We assume that Bienor is rotating around its pole in a prograde manner, with a period of 9.1736 h, and the resulting backplanes and radiance are sampled with a 5-degree step. The synthetic rotational light curve for Bienor as observed from Earth on December 26, 2022, is presented in Fig. 7. An animation depicting the rotation of Bienor, which generates this synthetic light curve, can be accessed through the following link.

We projected our ellipsoid onto the sky plane at the rotational phase value calculated for this occultation (0.15), with the north oriented upwards along the vertical axis and the east pointing to the left. Next, we also projected the chords onto the same frame and manually adjusted the angular size occupied by Bienor in that FoV to match the extremities of the chords described in Section 3.1 (see Fig. 8). The alignment agrees with that fitted ellipse in Fig. 2b.

Then, we replicated the process mentioned above with the chords from Occ. A and C (refer to Fig. C.1 of the Appendix B). We accounted for the geocentric distance ($\Delta$), being the shape model projection to scale for each case. Interestingly, we observe that the chord acquired from Van Vleck does not fit the projected shape. Since unlike Westport, Van Vleck lacked GPS synchronization, we assume this chord must be shifted. Upon subtracting 4 seconds from the ingress and egress times of this occultation, we obtain a chord that aligns with the ellipsoidal shape of our simulation (Fig. C.1b).

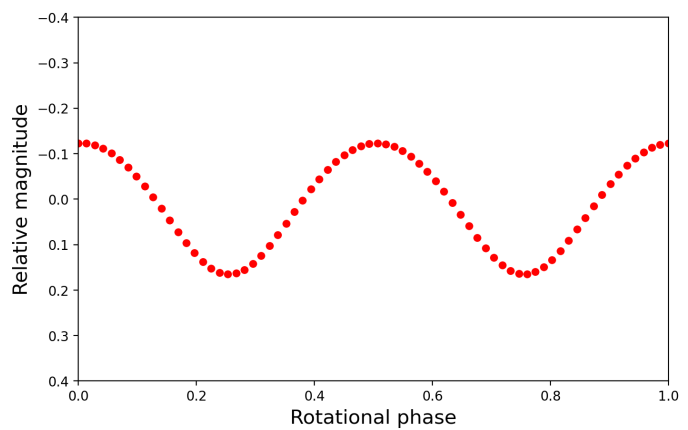

Fig. 7: Synthetic rotational light curve generated for Bienor, as observed from Earth on December 26, 2022. The light curve was sampled at intervals of 5 degrees.

With the new refined rotation period determined in this work (9.1736 ± 0.0002 h), we computed the rotational phase of Bienor on 11 January 2019, at 01:03:30.00 UT, obtaining a value of 0.67 ± 0.015, when the maximum of brightness is selected as phase 0. Then, we simulated Bienor's shape as seen from Earth (see Fig. C.3 of Appendix B). At this time, a stellar occultation by Bienor was analyzed by Fernández-Valenzuela et al. (2023), and we can see that our figure is again consistent with their projected shape (see Fig. 5 in that paper).

Finally, by applying Eq. 8 and 9, we compute the value of $\theta_v$ and compare with the chords obtained in Section 3.1. Remarkably, these values also align with our simulations (see Fig. C.2 in Appendix B).

Nevertheless, as shown in Fig. 7, this ellipsoid model cannot account for the asymmetries identified in the observational rotational light curves. The next section addresses several hypotheses that could explain the observational results.

## 5. Discussion

By observing three positive occultations from Japan, Eastern Europe, and the USA, we have determined the shape of the Bienor as a triaxial ellipsoid. Then, we retrieved the absolute axis of the ellipsoid and through the aspect angle and rotational phase, we theoretically computed the inclination of the minor axis of the projected ellipse onto the sky plane. This result together with the determined ellipsoid shape, allowed us to confirm that Bienor rotates in a prograde manner. Subsequently, synthetic light curves were generated, validating the rotational phases, size, shape, and sky-plane chords observed in all the stellar occultations by Bienor to date.

However, based on the assumption of a triaxial ellipsoid, our simulations predict symmetric rotational light curves. In contrast, the observed rotational curves display notable asymmetry, characterized by an absolute minimum following the absolute maximum, with a flux drop ~45% deeper than the relative one for 2023.

This discrepancy may arise from a variety of factors. A first alternative is that Bienor presents a flattened side making irreg-ular its shape but not detected in the occultations. To check this idea, we flattened one of Bienor's extremities in a way that this deformation is not visible when Bienor is in the configuration determined for Occ. B. Then, to make the simulation more realistic, we incorporate in our software the Minnaert photometric model (Minnaert 1941) to compute the radiance:

$$cos(i)^k cos(e)^{k-1} \qquad (10)$$

where $i$ and $e$ are the incidence and emission angles. We give the parameter $k$ a value of 2.5, so the model breaks the isotropy of reflected light, accounting for the facets' directionality relative to the observer.

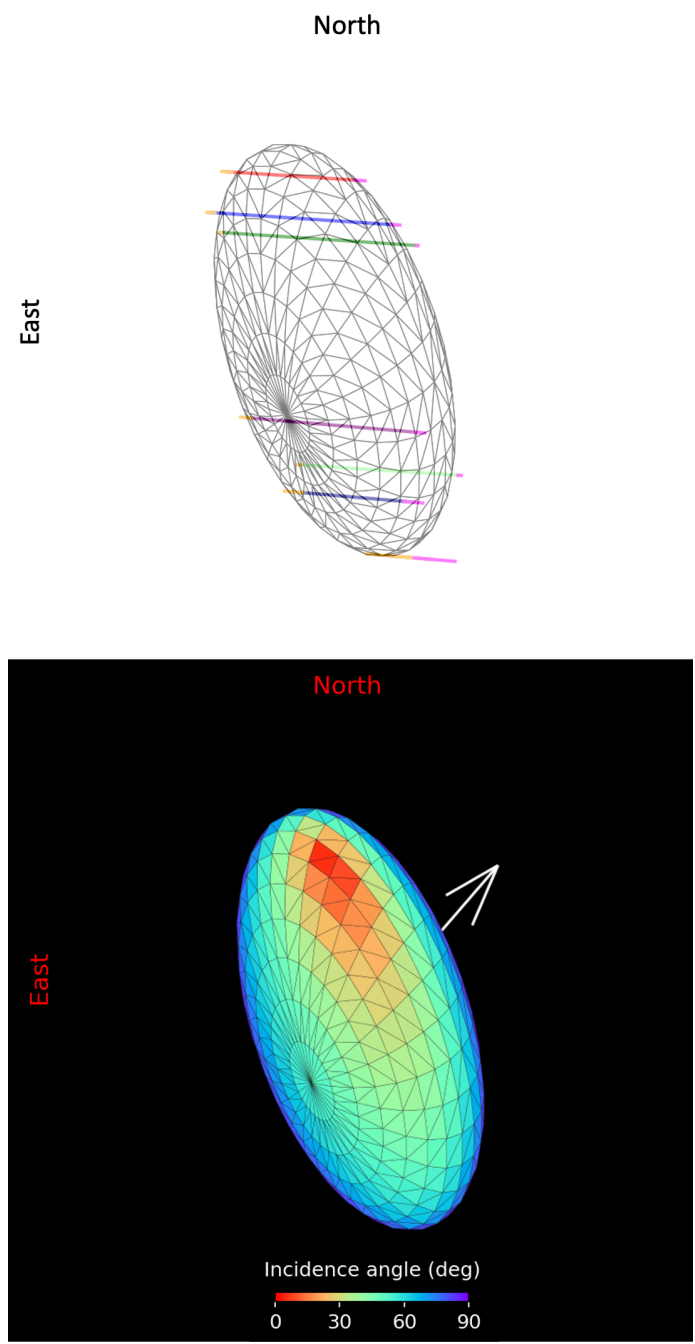

Fig. 8: Bienor shape model as observed from Earth during Occ. B at a rotational phase of 0.15. The top panel showcases chords from Occ. B overlaid onto the shape model, demonstrating that the shape of the projected ellipse corresponds closely to the least square-fitted ellipse outlined in Section 3.1. In the bottom panel, the same but with coloring. The colors denote the solar incidence angle. The white arrow points toward the rotation pole.

The resulting rotational light curve is asymmetrical (Fig. D.1a in Appendix C), but only with a difference of ~15% between minima.

Another different idea would be to assume that Bienor is a contact binary. This is not unreasonable given that to date, we know of the existence of contact binaries in the solar system such as the comets 67P/Churyumov-Gerasimenko (Sierks et al. 2015) or Tuttle (Harmon et al. 2010), the asteroids 4769 Castalia (Bottke & Melosh 1996) or 25143 Itokawa (Campo Bagatin et al. 2020), as well as trans-Neptunian objects like 486958 Arrokoth (Buie et al. 2020). These bodies are thought to have formed either through low-speed collisions during the early stages of the solar system or through gravitational re-accumulation following catastrophic collisions.

To explore this hypothesis, we generated a contact binary by distorting our triaxial ellipsoid. The resulting synthetic rotational light curve lacked asymmetry (Fig. D.1b). To eliminate the possibility that the highly symmetrical shape of this distorted shape model was responsible, we opted to append a smaller spherical body to the equator of the original ellipsoid, as illustrated in Fig. D.1c, thereby creating a markedly asymmetric body with a bulge at one end. However, again the resulting light curve remained lacking in asymmetry. Then, we introduced protuberancy at a different point along the equatorial region, which, when combined with the viewing geometry, disrupted the system's symmetry. In this scenario, the rotational light curve displayed asymmetry (Fig. D.1d), with less than a 20% difference between the minima.

In none of the above cases do we obtain values close to the ~45% measured in the real light curve. This, in addition to the absence of chords consistent with a binary system, and the inability to explain the size/albedo differences when comparing to thermal measurements, leads us to discard this idea as a sole explanation.

Another hypothesis for these asymmetries would be that certain regions of Bienor possess lower albedo. There are numerous cases of objects in the solar system, such as Earth's moon, Iapetus, the outermost of Saturn's large moons, or even Pluto, which exhibit clear differences in albedo on their surfaces. The asymmetries appear to intensify as Bienor approaches a reduced aspect angle (see Fig. D.2, spanning from 2016 to 2023), indicating that this low albedo region, if present, likely resides towards the equator, making it more prominent as it is more visible. To investigate this, we conducted a simulation of a light curve using the ellipsoid shape model. For this case, we modified the software to incorporate a region centered at the equator with a lower albedo, specifically, an albedo half that of the rest of the surface. As a result, a significant decrease in brightness (60%) is replicated (Fig. D.1e).

Finally, a last plausible explanation would be the presence of a secondary body, suggesting that the observed rotational light curve asymmetries are the result of mutual eclipses between a moon and the primary body. This rationale is further substantiated on the one hand by the fact that our photometric data in the visible range lead to an extrapolated area-equivalent diameter of 158 km, contrasting with the range of 179184 km determined by Lellouch et al. (2017). Additionally, while Bienor stands out as one of the least red centaurs, with a B-V value of 1.12 ś 0.03 (Tegler et al. 2008), and previous thermal measurements indicating a geometric albedo value of ~5.0% (Bauer et al. 2013; Duffard et al. 2014a; Lellouch et al. 2017), positioning it within the dark-neutral category, our analysis, reveals a higher geometric albedo of $(6.5 \pm 0.5)$%. If during the Occ. B, the area-equivalent diameter were 170 km (instead of the 149 km determined by ellipse-fit), the geometric albedo would be ∼5%. Therefore, this discrepancy in geometric albedo could be also explained by the presence of an additional object, potentially accounting for a missing reflecting area. A similar situation has been proposed for the trans-Neptunian object 2002 $TC_{302}$ (see Ortiz et al. 2020).

To explore this scenario, we introduced a moon (same albedo as Bienor), comprising 16% of the volume of the triaxial ellipsoid, positioned at a distance 1.6 times the value of the semi-major axis $a$, and situated at Bienor's equatorial plane. We performed the simulations for November 15, 2023, the date when the rotational light curve showed the greatest asymmetry. However, upon computing the reflected light across a grid of orbital phases with a 10-degree increment (defining orbital phases as the Moon-Earth-Bienor angle), the expected asymmetric light curve did not appear. This arises from the fact that the alignment of Bienor, the Sun, and the Earth does not yield mutual eclipses (primary and secondary) for a moon in an equatorial orbit under such configurations.

Then, we investigated this hypothesis by adjusting the inclination plane of the satellite in increments of 5 degrees, simulating light curves, and revisiting the outcomes. Certain configurations, due to mutual eclipse events, produced asymmetric light curves (as illustrated in Fig. D.1e, exhibiting an 18% asymmetry between minima). However, achieving such asymmetric behavior necessitates the rotational plane of the moon to possess an inclination of around 125° relative to Bienor's equatorial plane. Unfortunately, our available data is insufficient to determine the axes' dimension, orbital inclination, distance, or shape of this hypothetical moon. There would also be the possibility that it is not a solid object, but rather a ring, with a non-homogeneous density. Although no discernible drop in signal was observed in any occultation light curve, given that the signal-to-noise ratio (SNR) was not sufficiently high, it does not allow us to dismiss the possibility of an additional companion entirely.

The observed asymmetries and discrepancies in size estimations are sufficient evidence to support the notion of a satellite or ring's existence, while not excluding the possibility of irregularities on Bienor's surface or variations in albedo. However, there is no data that favors or completely discards any of the hypotheses. It is even plausible that both hypotheses are simultaneously valid. Further data collection is necessary to constrain and fully understand this system.

## 6. Conclusions

In this work, the analysis of three positive occultations of Bienor from Japan, Eastern Europe, and the United States, together with the rotational light curve data gathered from Spain, as well as the development of a tool for generating synthetic light curves, allowed us to:

1. Determine the projected ellipse of Bienor for December 26, 2022, showing that it is compatible with the proposed semi-axis ratios of $b/a = 0.45$ and $c/b = 0.79$.
2. Confirm a prograde rotation of Bienor and validate that the rotational pole solution of $\beta_p = 50°$, $\lambda_p = 35°$, aligns with observations spanning 22 years.

3. Confirm the presence of a clear asymmetry in the rotational light curve with distinct absolute and relative minima.
4. Refine the rotation period of Bienor to $9.1736 \pm 0.0002$ hours.
5. Determine the absolute values of the axes of the triaxial ellipsoid: a = $(127 \pm 5)$ km, b = $(55 \pm 4)$ km, and c = $(45 \pm 4)$ km. Using these values, we compare the area-equivalent diameter with thermal estimates and find that our solution yields a lower value.
6. Determine a geometric albedo for Bienor of $(6.5 \pm 0.5)$ %, higher than the geometric albedo determined previously by other methods based on thermal data.
7. Show through the three-dimensional simulation of rotational light curves that a combination of shape irregularities, the presence of an additional object, such as a satellite, and significant albedo differences on the surface, could explain the measured discrepancies.

Unfortunately, no occultation to date has recorded flux measurement to constrain the problem and discard any of the hypotheses presented here. In the future, new data will eventually clarify the issue.

*Acknowledgements.* J. L. Rizos, J.L. Ortiz, N. Morales, P. Santos-Sanz, M. Varia-Lubiano, R. Leiva, M. Kretlow, R. Morales, A. Alvarez-Candal, R. Duffard and J. M. Gómez-Limón acknowledge financial support from the Severo Ochoa grant CEX2021-001131-S funded by MCIN/AEI/10.13039/501100011033. J. L. Rizos acknowledges support from the Ministry of Science and Innovation under the funding of the European Union NextGeneration EU/PRTR. P. Santos-Sanz acknowledges financial support from the Spanish I+D+i project PID2022-139555NB-I00 (TNO-JWST) funded by MCIN/AEI/10.13039/501100011033. This work has made use of data from the European Space Agency (ESA) mission Gaia, processed by the Gaia Data Processing and Analysis Consortium (DPAC). Funding for the DPAC has been provided by national institutions, in particular the institutions participating in the Gaia Multilateral Agreement. This work is partly based on observations collected at the Centro Astronómico Hispano en Andalucía (CAHA) at Calar Alto, operated jointly by Junta de Andalucía and Consejo Superior de Investigaciones Científicas (CSIC). This research is also partially based on observations carried out at the Observatorio de Sierra Nevada (OSN) operated by Instituto de Astrofísica de Andalucía (IAA-CSIC).

## References

Bauer, J. M., Choi, Y.-J., Weissman, P. R., et al. 2008, PASP, 120, 393
Bauer, J. M., Grav, T., Blauvelt, E., et al. 2013, ApJ, 773, 22
Bottke, William F. J. & Melosh, H. J. 1996, Icarus, 124, 372
Braga-Ribas, F., Pereira, C. L., Sicardy, B., et al. 2023, A&A, 676, A72
Braga-Ribas, F., Sicardy, B., Ortiz, J. L., et al. 2014, Nature, 508, 72
Buie, M. W., Porter, S. B., Tamblyn, P., et al. 2020, AJ, 159, 130
Campo Bagatin, A., Alemañ, R. A., Benavidez, P. G., Pérez-Molina, M., & Richardson, D. C. 2020, Icarus, 339, 113603
Collins, K. A., Kielkopf, J. F., Stassun, K. G., & Hessman, F. V. 2017, AJ, 153, 77
DeMeo, F. E., Fornasier, S., Barucci, M. A., et al. 2009, A&A, 493, 283
Desmars, J., Camargo, J. I. B., Braga-Ribas, F., et al. 2015, A&A, 584, A96
Di Sisto, R. P. & Brunini, A. 2007, Icarus, 190, 224
Dotto, E., Barucci, M. A., Boehnhardt, H., et al. 2003, Icarus, 162, 408
Duffard, R., Pinilla-Alonso, N., Ortiz, J. L., et al. 2014a, A&A, 568, A79
Duffard, R., Pinilla-Alonso, N., Santos-Sanz, P., et al. 2014b, A&A, 564, A92
Duncan, M. J. & Levison, H. F. 1997, Science, 276, 1670
Duncan, M. J., Levison, H. F., & Budd, S. M. 1995, AJ, 110, 3073
Durech, J., Sidorin, V., & Kaasalainen, M. 2010, A&A, 513, A46
Elliot, J. L., Kern, S. D., Clancy, K. B., et al. 2005, AJ, 129, 1117
Fernández-Valenzuela, E., Morales, N., Vara-Lubiano, M., et al. 2023, A&A, 669, A112
Fernández-Valenzuela, E., Ortiz, J. L., Duffard, R., Morales, N., & Santos-Sanz, P. 2017, MNRAS, 466, 4147
Guilbert, A., Alvarez-Candal, A., Merlin, F., et al. 2009, Icarus, 201, 272
Halíř, R. & Flusser, J. 1998, in International Conference in Central Europe on Computer Graphics and Visualization
Harmon, J. K., Nolan, M. C., Giorgini, J. D., & Howell, E. S. 2010, Icarus, 207, 499
Harris, A. W., Young, J. W., Scaltriti, F., & Zappala, V. 1984, Icarus, 57, 251

Holman, M. J. & Wisdom, J. 1993, AJ, 105, 1987
Horner, J., Evans, N. W., & Bailey, M. E. 2004, Monthly Notices of the Royal Astronomical Society, 354, 798
Hunter, J. D. 2007, Computing in Science & Engineering, 9, 90
Kilic, Y., Braga-Ribas, F., Kaplan, M., et al. 2022, MNRAS, 515, 1346
Lacerda, P., Fornasier, S., Lellouch, E., et al. 2014, ApJ, 793, L2
Leiva, R., Sicardy, B., Camargo, J. I. B., et al. 2017, AJ, 154, 159
Lellouch, E., Moreno, R., Müller, T., et al. 2017, A&A, 608, A45
Lomb, N. R. 1976, Ap&SS, 39, 447
Magnusson, P. 1986, Icarus, 68, 1
Masiero, J. R., Wright, E. L., & Mainzer, A. K. 2021, The Planetary Science Journal, 2, 32
Minnaert, M. 1941, ApJ, 93, 403
Morgado, B. E., Sicardy, B., Braga-Ribas, F., et al. 2021, A&A, 652, A141
Müller, T., Lellouch, E., & Fornasier, S. 2020, in The Trans-Neptunian Solar System, ed. D. Prialnik, M. A. Barucci, & L. Young, 153–181
Ortiz, J. L., Baumont, S., Gutiérrez, P. J., & Roos-Serote, M. 2002, A&A, 388, 661
Ortiz, J. L., Duffard, R., Pinilla-Alonso, N., et al. 2015, A&A, 576, A18
Ortiz, J. L., Pereira, C. L., Sicardy, B., et al. 2023, A&A, 676, L12
Ortiz, J. L., Santos-Sanz, P., Sicardy, B., et al. 2020, A&A, 639, A134
Ostro, S. J., Connelly, R., & Dorogi, M. 1988, Icarus, 75, 30
Pereira, C. L., Braga-Ribas, F., Sicardy, B., et al. 2024, MNRAS, 527, 3624
Press, W. H., Teukolsky, S. A., Vetterling, W. T., & Flannery, B. P. 1992, Numerical recipes in FORTRAN. The art of scientific computing
Rousselot, P., Kryszczyńska, A., Bartczak, P., et al. 2021, MNRAS, 507, 3444
Russell, H. N. 1916, ApJ, 43, 173
Santos-Sanz, P., Ortiz, J. L., Sicardy, B., et al. 2021, MNRAS, 501, 6062
Showalter, M. R., Benecchi, S. D., Buie, M. W., et al. 2021, Icarus, 356, 114098
Sierks, H., Barbieri, C., Lamy, P. L., et al. 2015, Science, 347, aaa1044
Strauss, R. H., Leiva, R., Keller, J. M., et al. 2021, The Planetary Science Journal, 2, 22
Tegler, S. C., Bauer, J. M., Romanishin, W., & Peixinho, N. 2008, in The Solar System Beyond Neptune, ed. M. A. Barucci, H. Boehnhardt, D. P. Cruikshank, A. Morbidelli, & R. Dotson, 105–114
Tegler, S. C., Romanishin, W., Consolmagno, G. J., & J., S. 2016, AJ, 152, 210
van Belle, G. T. 1999, PASP, 111, 1515
Zacharias, N., Monet, D. G., Levine, S. E., et al. 2004, in American Astronomical Society Meeting Abstracts, Vol. 205, American Astronomical Society Meeting Abstracts, 48.15

[1] Instituto de Astrofísica de Andalucía Consejo Superior de Investigaciones Científicas (IAA-CSIC), Glorieta de la Astronomía S/N, E-18008, Granada, Spain. e-mail: jlrizos@iaa.es
[2] Florida Space Institute, UCF, 12354 Research Parkway, Partnership 1 building, Room 211, Orlado, USA
[3] Federal University of Technology-Paraná (UTFPR/DAFIS), Av. Sete de Setembro, 3165, CEP 80230-901 - Curitiba - PR - Brazil
[4] Laboratório Interinstitucional de e-Astronomia - LIneA - and INCT do e-Universo. Av. Pastor Martin Luther King Jr, 126 Del Castilho, Nova América Offices, Torre 3000 / sala 817 CEP: 20765-000, Brazil
[5] LESIA, UMR8109, Observatoire de Paris, Université PSL, CNRS, Sorbonne Université, Universitde Paris, 5 place Jules Janssen, Meudon, 92195, France
[6] Instituto de Física Aplicada a las Ciencias y las Tecnologías, Universidad de Alicante, San Vicent del Raspeig, E03080, Alicante, Spain
[7] Space Telescope Science Institute, 3700 San Martin Drive, Baltimore, USA
[8] Institut Polytechnique des Sciences Avancées IPSA, 63 boulevard de Brandebourg, F-94200 Ivry-sur-Seine, France
[9] Institut de Mécanique Céleste et de Calcul des Éphémérides, IMCCE, Observatoire de Paris, PSL Research University. CNRS,Sorbonne Universités, UPMC Univ Paris 06, Univ. Lille, 77 Av. Denfert-Rochereau, F-75014 Paris, France
[10] naXys, Department of Mathematics, University of Namur, Rue de Bruxelles 61, Namur, 5000, Belgium
[11] Universidade Federal do Rio de Janeiro, Observatório do Valongo, Ladeira Pedro do Antonio 43, Rio de Janeiro, RJ 20.080-90, Brazil
[12] UNESP - São Paulo State University, Grupo de Dinâmica Orbital e Planetologia, Guaratinguetá, SP, 12516-410, Brazil
[13] Observatório Nacional (MCTI), Rua Gal. José Cristino, 77Bairro Imperial de São Cristóvão, 20921-400 Rio de Janeiro, Brazil
[14] TÜBTAK National Observatory, Akdeniz University Campus, 07058 Antalya, Turkey
[15] Astronomy Department and Van Vleck Observatory, Wesleyan University, Middletown, CT 06459, USA
[16] Westport Astronomical Society, Westport, CT, USA
[17] Dept of Mathematics and Physics, University of New Haven, West Haven, CT USA
[18] Salvia Observatory, France
[19] Schreurs O (S.A.L.), Liège, Belgium
[20] Société Astronomique de Liège (S.A.L), Nandrin Observatory, Belgium
[21] Dourbes, Viroinval, Belgium
[22] Tsu Mie, Tsu Mie, Japan
[23] Pressigny, France
[24] Uda Nara, Nara, Japan
[25] Owase Mie, Nakamuracho, Japan
[26] Cottered Observatory, Hertfordshire, England

# Appendix A: Occulted stars

Hertzsprung–Russell diagram showing the absolute magnitude versus the color index (B–V) of stars from the HYG star database archive[6], which combines data from the HIPPARCOS, Yale Bright Star, and Gliese (nearby star) catalogs. The occulted stars analyzed in Section 2 are plotted, including Occ. A (202357903847210880 - GaiaDR3) and Occ. B (961233167113907328 - GaiaDR3), which are giant stars, and Occ. C (193308850132660224 - GaiaDR3), which is a main-sequence star.

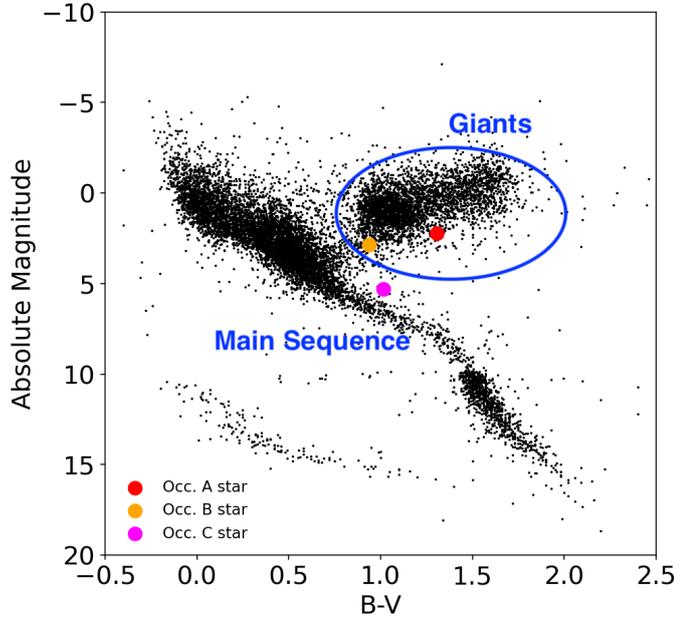

Fig. A.1: Absolute magnitude versus color index (B–V) of the HYG star database, along with the stars occulted by Bienor and analyzed in this work.

# Appendix B: Stellar occultation analyses

Light curves from the occultation events described in Section 3.1. Additionally, tables containing the ingress and egress times after the square well fits are provided, along with the expanded table containing information about the location of telescopes, instruments, and observers.

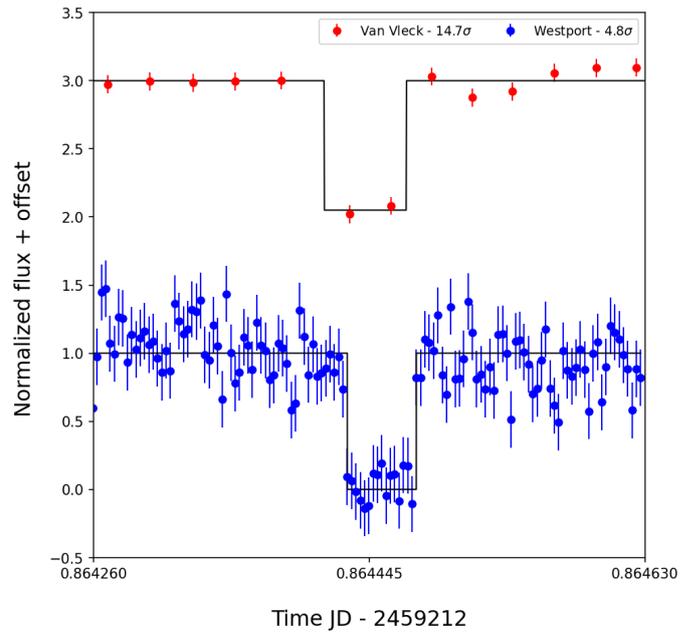

Fig. B.1: Normalized flux value (with offset for better visualization) obtained from the two observations made during Occ. A on February 6, 2022. They present a flux drop of 17.4$\sigma$ for Van Vleck (red) and 4.8$\sigma$ for Westport (blue). The black line represents a square-well fit to the observational data.

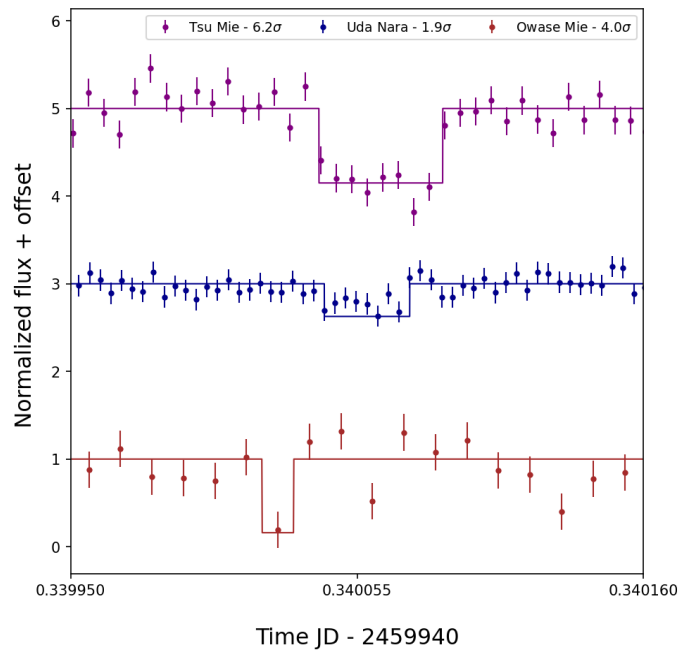

Fig. B.2: Normalized flux value (with offset for better visualization) obtained from the three observations made during Occ. B from Japan on December 26, 2022. The flux drop (measured in $\sigma$) appears at the top. The black line represents a square-well fit to the observational data.

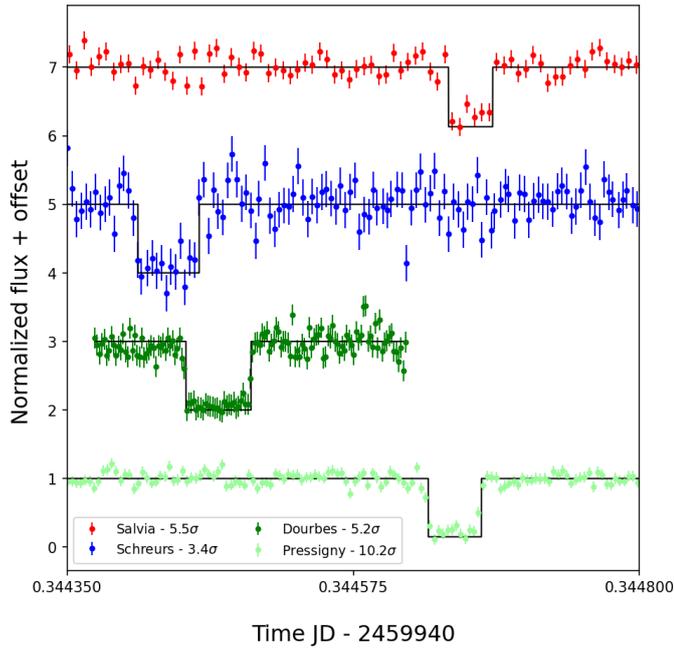

Fig. B.3: Normalized flux value (with offset for better visualization) obtained from the four observations made during Occ. B from Western Europe on December 26, 2022. The flux drop (measured in $\sigma$) appears at the top. The black line represents a square-well fit to the observational data.

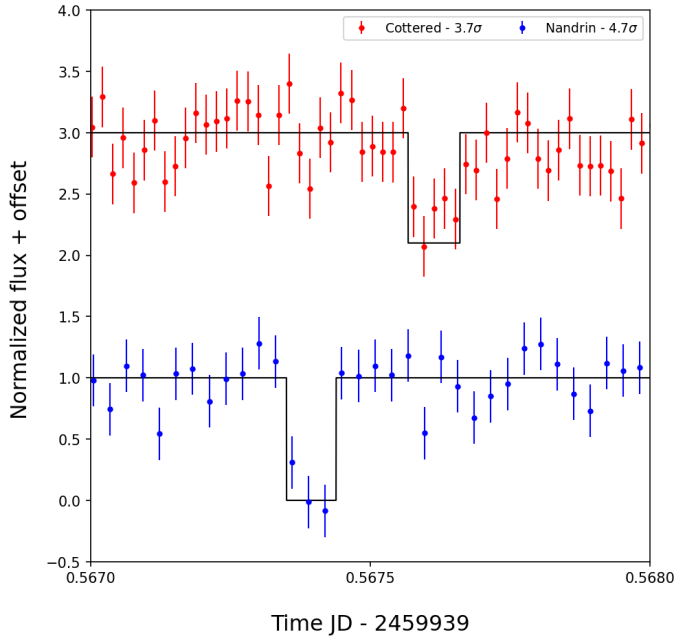

Fig. B.4: Normalized flux value (with offset for better visualization) obtained from the two observations made during Occ. C, on February 14, 2024. They present a flux drop of $3.7\sigma$ for Cottered (red) and $4.7\sigma$ for Nandrin S.A.L (blue). The black line represents a square-well fit to the observational data.

Table B.1: Observatories, instrumentation, and observers.

| Date | Observatory | Longitude (°) | Latitude (°) | Altitute (m) | Telescope | Detector | Exposure time (s) | Cicle time (s) | Synchronization | Occultation | Observers |
|---|---|---|---|---|---|---|---|---|---|---|---|
| 6 Feb, 2022 | Van Vleck, USA | -72.65 | 41.55 | 68.9 | 610 mm - f/0.15 | Apogee E2V CCD42-40 | 0.800 | 2.320 | NTP | Positive | Seth Redfield Cassidy Soloff Kyle McGregor |
| 6 Feb, 2022 | Westport Astronomical Society, USA | -73.32 | 41.17 | 88.0 | 356 mm - f/0.12 | QHY174M-GPS | 0.250 | 0.250 | GPS | Positive | Kevin Green |
| 26 Dic, 2022 | Salvia, France | -0.40 | 47.98 | 99.0 | 356 mm - f/3.9 | QHY174M-GPS | 0.500 | 0.500 | GPS | Positive | Thierry Midavaine |
| 26 Dic, 2022 | Schreurs Liège, Belgium | 5.55 | 50.64 | 107.0 | 254 mm - f/0.21 | Watec 910 HX /RC | 0.320 | 0.320 | GPS | Positive | Olivier Schreurs |
| 26 Dic, 2022 | Dourbes, Belgium | 4.58 | 50.09 | 192.0 | 400 mm - f/0.22 | Watec 910HX | 0.160 | 0.160 | GPS | Positive | Roland Boninsegna |
| 26 Dic, 2022 | Tsu Mie, Japan | 136.45 | 34.66 | 5.4 | 200 mm - f/0.10 | ZWO ASI290MM | 0.490 | 0.490 | GPS | Positive | Miyoshi Ida |
| 26 Dic, 2022 | Le Grand Pressigny, France | 0.78 | 46.92 | 58.0 | 500 mm - f/0.30 | QHY174M-GPS | 0.300 | 0.300 | GPS | Positive | Pierre Le Cam |
| 26 Dic, 2022 | Uda Nara, Japan | 136.01 | 34.57 | 407.0 | 200 mm - f/0.10 | ZWO ASI290MM | 0.338 | 0.338 | GPS | Positive | Ken Isobe |
| 26 Dic, 2022 | Owase Mie, Japan | 136.19 | 34.07 | 37.0 | 220 mm - f/0.10 | ZWO ASI290MM | 0.499 | 0.499 | GPS | Positive | Hayato Watanabe Syouji Yuasa Hikaru Watanabe |
| 14 Feb, 2023 | Cottered, England | -0.06 | 51.95 | 120.0 | 350 mm - f/0.22 | ZWO ASI174MM | 1.603 | 1.603 | GPS | Positive | Simon Kidd |
| 14 Feb, 2023 | Nandrin S.A.L, Belgium | 5.44 | 50.52 | 261.0 | 406 mm - f/0.35 | Watec 910 HX /RC | 2.56 | 2.56 | GPS | Positive | Olivier Schreurs Manon Lecossois |

| Chord | Ingress time | Egress time |
|-------|-------------|-------------|
| Occ. A - 06 Feb. 2022 | | |
| Van Vleck | 08:44:44.4800* ± 1.5600 | 08:44:51.6700* ± 1.5600 |
| Westport | 08:44:46.7731 ± 0.0461 | 08:44:50.7746 ± 0.0461 |
| Occ. B - 26 Dec. 2022 | | |
| Salvia | 20:16:18.0034 ± 0.1078 | 20:16:21.0038 ± 0.1078 |
| Schreurs | 20:15:56.6541 ± 0.0883 | 20:16:00.8141 ± 0.0883 |
| Dourbes | 20:15:59.9112 ± 0.0174 | 20:16:04.3342 ± 0.0174 |
| Tsu Mie | 20:09:39.6022 ± 0.1217 | 20:09:43.5234 ± 0.1217 |
| Pressigny | 20:16:16.3726 ± 0.0356 | 20:16:19.9788 ± 0.0356 |
| Uda Nara | 20:09:39.7147 ± 0.2112 | 20:09:42.4180 ± 0.2112 |
| Owase Mie | 20:09:37.7470 ± 0.4980 | 20:09:38.7440 ± 0.4980 |
| Occ. C - 14 Feb. 2023 | | |
| Cottered | 01:37:18.7117 ± 0.8173 | 01:37:26.7260 ± 0.8173 |
| Nandrin S.A.L | 01:36:59.9112 ± 0.5737 | 01:37:07.5912 ± 0.5737 |

Table B.2: UT ingress and egress times and errors for each observation. The Van Vleck times (marked with asterisks) were shifted by 4 seconds in Fig. C.1 to fit the projected shape model.

## Appendix C: Compatibility of chords with ellipsoidal

In this Appendix, we include the figures generated to study the compatibility of the projected chords in the sky plane with the orientation and size of the ellipsoid generated by our synthetic rotational light curve simulation software. Specifically, we include the figures of the chords superimposed on the projections of the shape model described in Section 4. In addition, we include diagrams showing the calculated position angle of the minor axis of the ellipse (Eq. 8 and 9) for Occ. A and C.

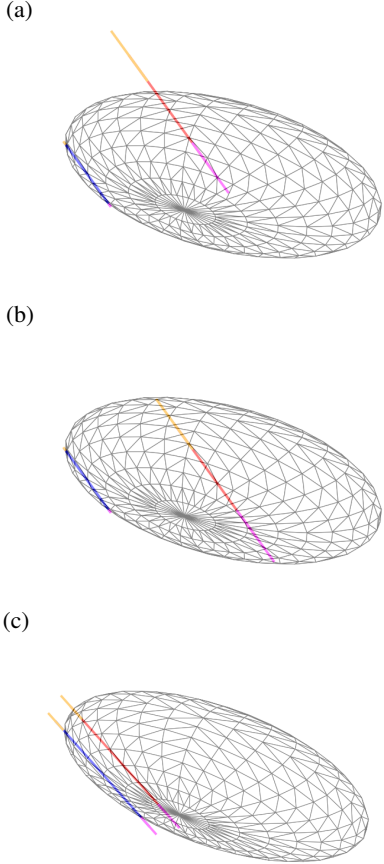

Fig. C.1: a) Chords from Occ. A superimposed onto the shape model, as seen from Earth (the north pointing upward and the east to the left), at a rotational phase value of 0.888. b) Same as on top but with Van Vleck's chord shifted 4.2 seconds so that it matches the projected shape (see Section 4.2). c) Chords from Occ. C overlaid onto the shape model as seen from Earth with a rotational phase of 0.93. Geocentric distance (Δ) is taken into account so that the size of the ellipsoids are to scale.

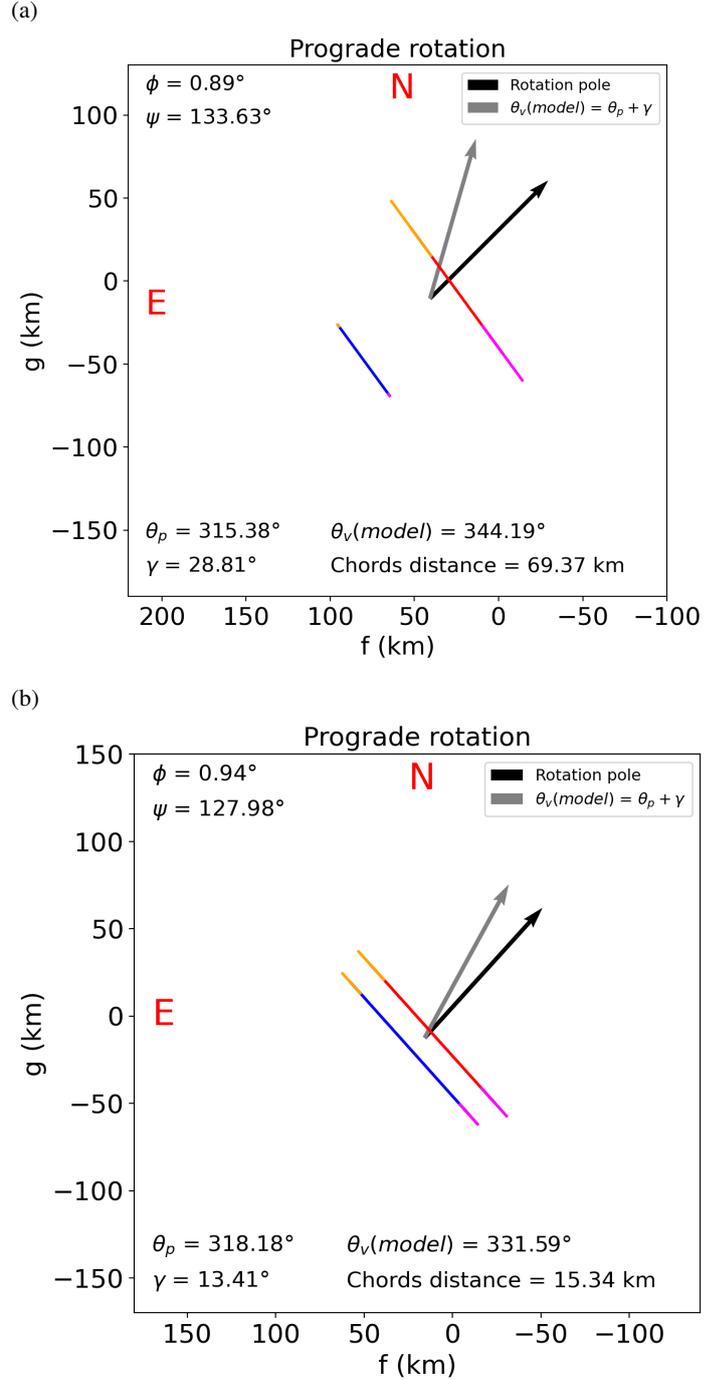

Fig. C.2: a) Occ. A's chords and value of the minor axis of the ellipse, determined using Eq. 8 and 9 for the prograde rotation. b) Same for Occ. C. Both solutions are consistent with the projected ellipse in Fig. C.1. To the right of each plot is included the position angle of the rotation pole ($\theta_p$), and minor axis of the projected ellipse ($\theta_v$), as well as the rotational phase and aspect angles.

(a)

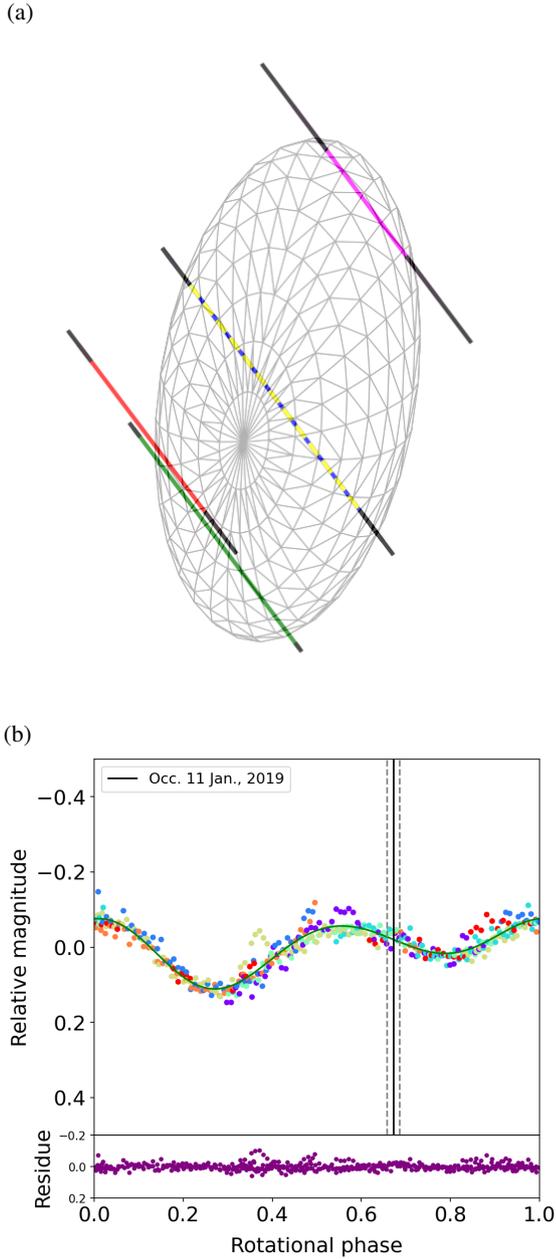

(b)

Fig. C.3: a) Shape model of Bienor, as seen from Earth on January 11, 2019, at 01:03:30 UT, with the chords of the occultation observed at that time and analyzed by Fernández-Valenzuela et al. (2023) (see Fig. 5 in their work). b) Rotational light curve using observational data from 2019 as reported by Fernández-Valenzuela et al. (2023). With the newly refined period from this work, the rotational phase was 0.67 ± 0.015. Dashed lines show the uncertainty after studying how the 0.0002 h error propagates with time. A unique color represents each observation day. At the bottom, it is shown the difference (residual) between the modeled curve and the observational data.

## Appendix D: Synthetic rotational light curves

In this appendix, the light curves generated through simulation software and the projections of the shape models used in this work are shown.

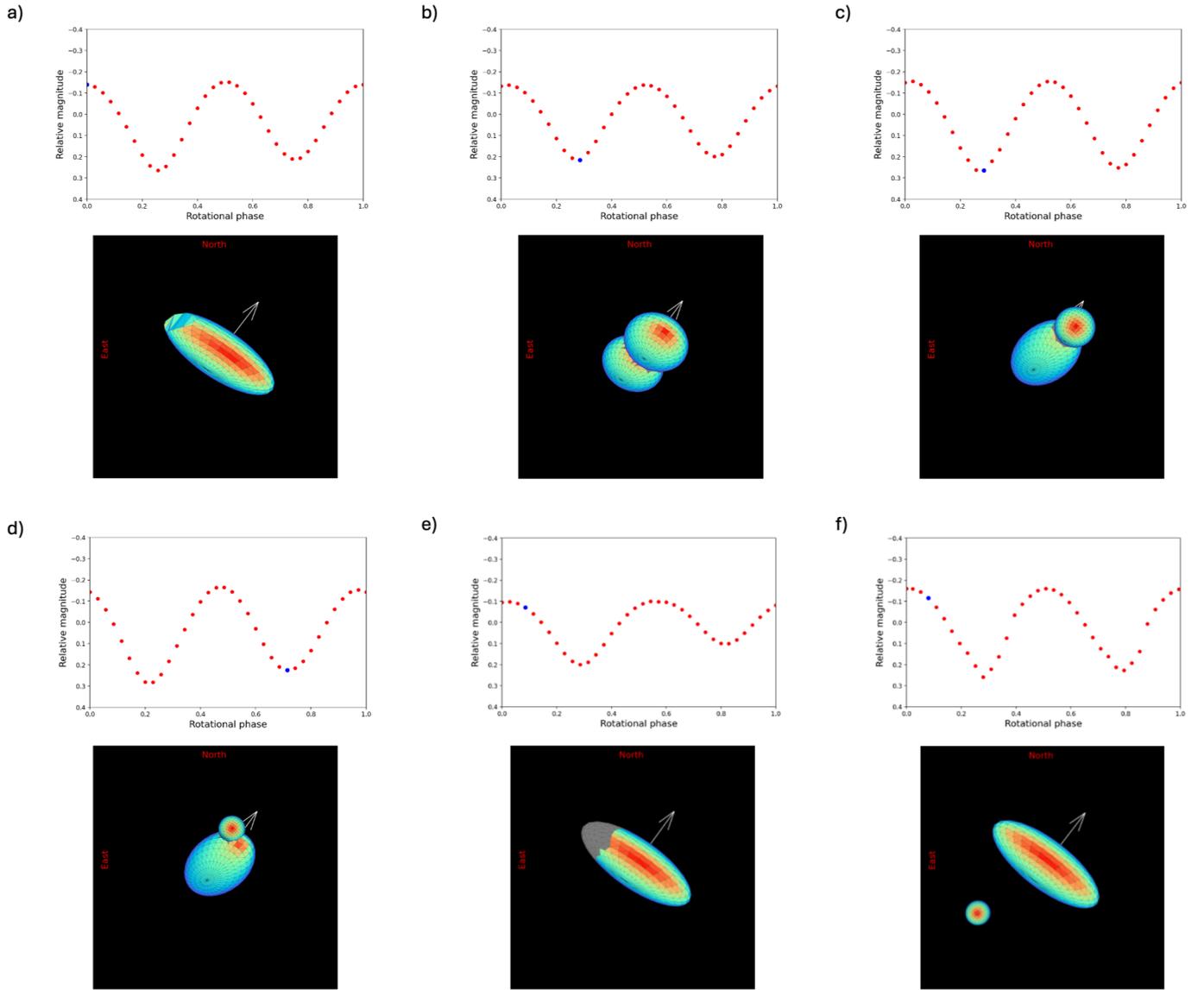

Fig. D.1: Simulated rotational light curves for various configurations observed from Earth on November 15, 2023. a) LC for the determined ellipsoidal shape for Bienor, but with a flattening at one of its ends. This configuration generates an asymmetric light curve with a ∼15% drop difference between minima. b) LC for a symmetric contact binary matching the dimensions of the ellipsoid shape model. No asymmetric light curve is generated. c) LC for an asymmetric contact binary, in which one of the lobes is smaller than the other. No asymmetric light curve is generated. d) Similar to (c), where the smaller object is now shifted from the symmetry axis of the ellipsoid. In this case, the minima are different, with a ∼20% difference in the drop. e) LC for an ellipsoid featuring a low-albedo region (half the regular albedo) in its equatorial area. In this example, up to 60% asymmetry between the minima is achieved. f) LC for an ellipsoid with a satellite of 16 % of its volume. This configuration, in which the orbital plane of the moon is inclined at 125° relative to the equatorial plane of Bienor (producing mutual eclipses), gives an asymmetric rotational with a difference of 18% drop between minima. An animation of each checked hypothesis can be accessed through the links a, b, c, d, e, and f. The white arrow points toward the rotation pole. The blue dot in the rotational light curve plots indicates the point corresponding to the image below. The colors indicate the value of the solar incidence angle, following the same scale shown in Fig. 6.

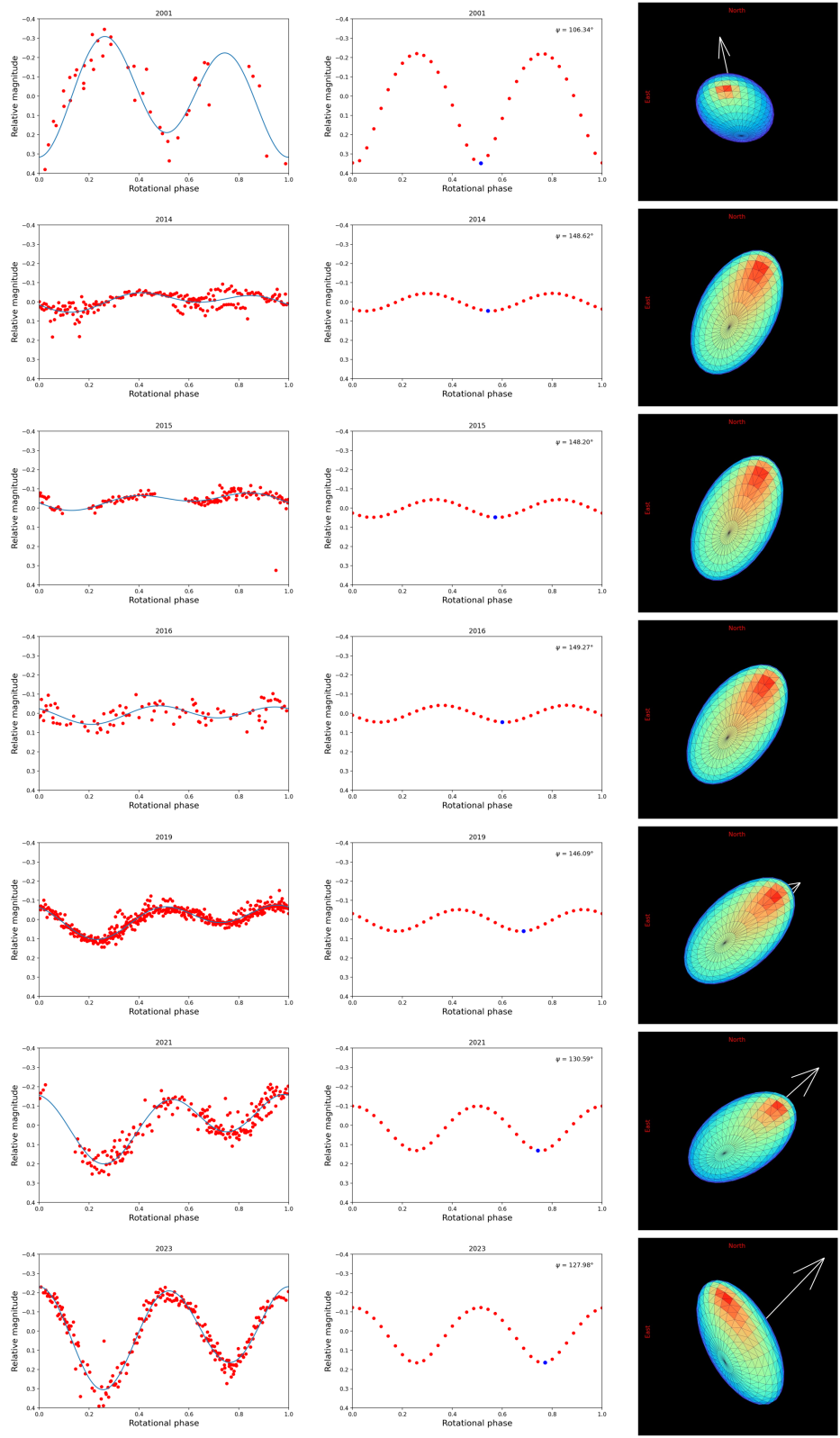

Fig. D.2: Left column: rotational light curves observed since 2001 (Ortiz et al. 2002; Fernández-Valenzuela et al. 2017, 2023). Julian date has been corrected for light travel time and phased using the refined period of 9.1736 h. Middle column: synthetic rotational light curves for a triaxial ellipsoid ($b/a = 0.45$, $c/b = 0.79$) using the prograde sense of rotation (Fernández-Valenzuela et al. 2017) generated by our software. At the upper right corner of each plot, the value of the aspect angle is displayed. The resulting backplanes and radiance are sampled with a 10-degree step. Right column: shape model as seen from Earth at the minimum brightness of the synthetic rotational light curve. The colors indicate the value of the solar incidence angle, following the same scale shown in Fig. 6. The white arrow points toward the rotation pole. The blue dot in the middle plots indicates the point that corresponds to the image on the right.